\documentclass[conference]{IEEEtran}
\IEEEoverridecommandlockouts

\usepackage{cite}
\usepackage{amsmath,amssymb,amsfonts}
\usepackage{amsthm}
\usepackage{algorithmic}
\usepackage{graphicx}
\usepackage{textcomp}
\usepackage{xcolor}
\def\BibTeX{{\rm B\kern-.05em{\sc i\kern-.025em b}\kern-.08em
    T\kern-.1667em\lower.7ex\hbox{E}\kern-.125emX}}
\begin{document}

\theoremstyle{remark}
\newtheorem*{remark}{Remark}

\allowdisplaybreaks

\title{\huge{Event-Triggered Distributed Target Tracking via PRIMEX}
\thanks{This work was supported by the National Natural Science Foundation of
China under grant No. 62501384.}
}

\author{\IEEEauthorblockN{Yuxuan Xia$^\star$, Kuo-Chu Chang$^\S$, Xueqi Qiu$^\P$, Lin Gao$^\dag$, Chaoqun Yang$^\ddag$, Ting Yuan$^\star$}
\IEEEauthorblockA{\footnotesize {$^\star${School of Automation and Intelligent Sensing},
{Shanghai Jiao Tong University},
Shanghai, China} \\
$^\S${Department of Systems Engineering and Operations Research},
{George Mason University},
Fairfax, VA, USA \\
$^\P${Department of Computer Science}, {Durham University}, Durham, the United Kingdom \\
$^\dag${University of Electronic Science and Technology of China}, Chengdu, China\\
$^\ddag${School of Automation}, {Southeast University}, Nanjing, China\\
Email: $^\star$yuxuan.xia@sjtu.edu.cn, $^\S$kchang@gmu.edu, $^\P$xueqi.qiu@durham.ac.uk, $^\dag$lin.gao@uestc.edu.cn, $^\ddag$ycq@seu.edu.cn, $^\star$tyuan@sjtu.edu.cn.}
}
\maketitle

\begin{abstract}
PRIMEX (prime-based graph encoding and extraction) is a recently proposed framework for scalable distributed fusion. In PRIMEX, the information pedigree of state estimates or probability density functions is encoded using the information codes, enabling lightweight arithmetic for redundancy removal and data integration. Building on PRIMEX and its memoryless fusion strategy based on a least-squares approximation, in this paper we present two efficient distributed tracking algorithms: a consensus-based PRIMEX method that fuses information from all neighbors, and a greedy gossip-based PRIMEX method that fuses with the most informative neighbor. To further increase communication efficiency, we incorporate an event-triggered mechanism, in which transmission decisions are driven by information novelty measured using differences between the information codes. The proposed methods are evaluated and compared with covariance intersection and centralized fusion in a distributed single
target tracking scenario. Simulation results show that PRIMEX-based methods remain competitive in tracking accuracy while
improving communication efficiency.

\end{abstract}


\section{Introduction}

Multi-sensor fusion has become a key enabler for modern perception and decision-making systems, including, e.g., autonomous driving, surveillance, and industrial remote monitoring \cite{xiang2023multi,liu2022multi}. By combining information from multiple sensors, fusion can improve estimation accuracy, robustness to failures, and resilience against model mismatch. 

A conventional way to implement sensor fusion is centralized fusion \cite{bar1995multitarget}. In a centralized architecture, all sensor measurements are transmitted to a fusion center, which maintains a global estimate using, e.g., a Kalman filter (in information form). This approach is conceptually simple and often yields near-optimal performance when communication is reliable and the network is small. However, centralized fusion suffers from several limitations as the communication costs grow with the number of sensors and the data dimension, and the architecture does not scale gracefully to large sensor networks.

These limitations motivate distributed fusion \cite{chong2003sensor}, where each node maintains a local estimate or probability density function and communicates only with its neighbors in a sensor network. Distributed fusion can improve scalability and robustness, and it allows the network to operate without a central coordinator. A key challenge in distributed fusion is managing redundant or inconsistently propagated information, which can degrade accuracy and increase decision-making risks. Methods based on information graphs (IGs) can provide structured fusion and address redundancy, but they require extensive communication and synchronization to maintain a consistent global view of information flow, making them impractical in large, dynamic sensor networks \cite{chong1990distributed}.

To address these challenges, PRIMEX, a prime-based graph encoding and extraction framework for efficient and scalable distributed fusion has recently been proposed \cite{chang2025primex}. In PRIMEX, instead of explicitly tracking the IG structures, the information pedigree of each state estimate or probability density function is encoded as a product of unique prime numbers based on the prime factorization theorem, where each prime represents a distinct information source. This enables fusion operations to leverage fundamental arithmetic --  the greatest common divisor (GCD) for identifying and removing redundancy and the least common multiple (LCM) for integrating new information. By doing so, we can achieve efficient distributed fusion with substantially lower communication and computational overhead. Unlike covariance intersection-based fusion methods that rely on conservative weightings to manage unknown correlations, PRIMEX explicitly encodes and algebraically removes shared information through information pedigree tracking, enabling redundancy removal  without sacrificing estimation accuracy.

PRIMEX achieves optimal fusion for static systems but provides only approximate fusion solutions for dynamical systems in the presence of common process noises \cite{chong1990distributed}. Moreover, exact PRIMEX implementations may require access to past fusion results, which can be inefficient in dynamical settings because stored components must be propagated forward in time. For distributed target tracking, it is therefore more practical to adopt a memoryless fusion strategy \cite{chang2025primex}. In addition, when retrieving the exact shared information becomes impractical due to network or storage constraints, PRIMEX performs fusion by selecting a set of locally available information that best approximates the required shared information. Mathematically, this leads to a constrained linear problem that can be solved using the least-squares approximation, yielding a solution that resembles covariance intersection \cite{chang2025primex,chang2025primex_isif}.

In the context of distributed tracking, there are additional design choices to be made. One key question is how each node aggregates information from its neighbors within each communication round. A natural solution is given by consensus-based fusion, where every node fuses with all neighbors that transmit in that round. The consensus-based fusion approach can propagate information quickly across the network, but may also result in redundant fusions and increased communication load, especially in dense connected networks.

An alternative is gossip-based fusion, where each node fuses with one neighbor, selected at random, per communication round \cite{kar2010gossip,ma2016gossip}. Gossip-based fusion reduces the computational burden at each iteration, but its convergence can be slow due to the randomized neighbor selection. In order to improve the convergence rate of the gossip process, greedy gossip strategies have been proposed \cite{wan2018distributed,shin2020sample}, in which, at each iteration, a node evaluates its neighbors and selects the one that can provide the most informative update for fusion. This, however, requires each node to evaluate all of its neighbors, which may increase the computational load. Therefore, for greedy gossip-based fusion, it is very important to design neighbor selection strategies that prioritize the most informative neighbors with minimum information exchange. This requirement aligns well with the prime-based information pedigree representation in PRIMEX. In PRIMEX, the new information that one node can provide to another is encoded in their prime-based information code, enabling efficient identification of the most informative neighbor for fusion.

Beyond the choice of consensus vs.\ gossip, communication efficiency can be further improved by making communication event-triggered rather than periodic or always-on. Instead of transmitting information to neighbors in every communication round, a node may choose to transmit information only when its local information has changed sufficiently as compared to its most recent transmission \cite{ge2019distributed}. In this paper, we also employ such an event-triggered mechanism where the decision is based on information novelty measured using differences between the prime-based information codes. This design allows us to exploit structural changes in the information pedigree when deciding whether communication is worthwhile without tracking the numerical changes in the local densities.

Motivated by these considerations, in this paper, we combine PRIMEX with consensus-based and greedy gossip-based fusion, along with event-triggered communication, to develop communication-efficient distributed tracking algorithms. The main contributions are summarized as follows:
\begin{itemize}
    \item We present two distributed tracking algorithms based on PRIMEX: a consensus-based PRIMEX method that fuses with all neighbors, and a greedy gossip-based PRIMEX method that fuses with the most informative neighbor.
    \item We incorporate an event-triggered communication mechanism in PRIMEX, enabling nodes to transmit only when their local information has changed sufficiently.
    \item We conduct a thorough simulation study on distributed single target tracking, comparing the proposed PRIMEX-based methods against covariance intersection and centralized fusion in terms of estimation accuracy and communication efficiency.
    \item Our simulation results show that PRIMEX-based methods achieve competitive tracking accuracy while substantially reducing communication overhead.
\end{itemize}

The remainder of the paper is organized as follows. Section II reviews optimal fusion within PRIMEX. Section III introduces problem formulation. Section IV presents the proposed consensus-based and gossip-based PRIMEX fusion algorithms with event-triggered communications. Section V presents the simulation results, and lastly Section VI concludes the paper and points out future research directions.

\section{Background}

In this section, we first review optimal fusion and then describe how it can be achieved within the PRIMEX framework.

\subsection{Optimal Fusion}

In a distributed sensor network, the sensors collect measurements $z^i$, $i = 1,\dots,N$, that provide information about an unknown state variable $x$. The goal of distributed fusion is to combine these measurements to generate an estimate or probability density function (pdf) of the state $x$. The fundamental principle for optimal fusion is Bayesian inference, where the posterior density of the state $x$ given all the measurements is
\begin{equation}
p(x \mid Z) = \frac{p(Z \mid x)p(x)}{\int p(Z \mid x)p(x) dx} \propto p(Z \mid x)p(x),
\end{equation}
where $Z = \{z_1,\dots,z_N\}$ is the set of all measurements, $p(x)$ is the prior density of $x$, and $p(Z \mid x)$ is the likelihood function of the measurements given $x$. Assuming that the measurements are conditionally independent given $x$, we have
\begin{equation} \textstyle 
p(x \mid Z) \propto \prod_{i=1}^N p(z^i \mid x)p(x),
\end{equation}
where $p(z^i \mid x)$ is the single measurement likelihood function, encoding the information provided by $z^i$.

To achieve optimal fusion, traditional approaches are developed using IGs \cite{chong1990distributed}. In an IG , nodes correspond to information sources or processors, whereas edges capture information and dependencies among nodes. Specifically, each node $i$ in an IG maintains a local density of $x$, and this density is updated by incorporating received information from a neighboring node $j$ according to
\begin{equation}
    p(x \mid Z_i \cup Z_j) \propto p(Z_i \cup Z_j \mid x)p(x), \label{eq_fused_info}
\end{equation}
where $Z_i \cup Z_j$ represents the fused information. In distributed fusion, overlapping information from multiple communication paths can result in biased estimates. This is addressed in IG-based fusion by decomposing \eqref{eq_fused_info} as \cite{chong1990distributed}
\begin{equation}
    p(x \mid Z_i \cup Z_j) \propto \frac{p(Z_i \mid x)p(Z_j \mid x)p(x)}{p(Z_i \cap Z_j \mid x)} \propto \frac{p(x \mid Z_i)p(x \mid Z_j)}{p(x \mid Z_i \cap Z_j)}, \label{eq_ig_fusion}
\end{equation}
where $Z_i \cap Z_j$ represents the shared (redundant) information between nodes $i$ and $j$. The denominator in \eqref{eq_ig_fusion} removes the effect of double counting the shared information during fusion.

When combining the information from multiple sources, the fusion of $n$ local densities $p(x \mid Z_i)$ can be represented as \cite{chong1990distributed}
\begin{equation} \textstyle 
    p(x \mid \cup_{i=1}^n Z_i) \propto \prod_{i=1}^n S_i^{(-1)^{i-1}}, \label{eq_multi_fusion}
\end{equation}
where $S_i$ represents the combination of $i$ local densities $p(x \mid Z_1),\dots, p(x \mid Z_i)$ such that
\begin{subequations}
    \begin{align}
    S_1 &= \textstyle \prod_{j=1}^n p(x \mid Z_j), \\
    S_2 &= \textstyle \prod_{1 \leq j_1 < j_2 \leq n} p(x \mid Z_{j_1} \cap Z_{j_2}), \cdots \\
    S_n &= \textstyle p(x \mid \cap_{j=1}^n Z_j). 
    \end{align}
\end{subequations}

In \eqref{eq_multi_fusion}, the alternating multiplication and division of densities effectively eliminates the conditional dependencies in the data by accounting for shared information. However, this requires the recognizing of correlated information resulting from past communication and fusion events and correctly identifying its presence in incoming data streams, making it impractical in distributed sensor networks with limited bandwidth \cite{chong1990distributed}.

\subsection{The PRIMEX Framework}

In IG-based approaches, performing data fusion according to \eqref{eq_multi_fusion} requires the maintenance of extensive pedigree information, and this limitation is mitigated in a recently proposed fusion framework PRIMEX \cite{chang2025primex}. Instead of explicitly tracking pedigrees, PRIMEX encodes them using prime factorization, enabling efficient fusion through simple arithmetic operations.

Specifically, in PRIMEX the information pedigree of node $i$ can be identified as (also called information code (IC) \cite{chang2025primex}) $I_i = p_1p_2\cdots p_\ell$, where $p_1,\dots,p_\ell$ are distinct prime numbers, each representing a unique information source. When a node receives information from another node, it can perform fusion by computing the LCM of the ICs corresponding to its current and received pdfs, effectively retaining all unique information while eliminating any redundant information that was previously received. For the shared information between two nodes, it can be identified by computing the GCD of their ICs. 

The above fusion process can be mathematically expressed as follows. Suppose that node $A$ with IC $I_A$ receives information from node $B$ with IC $I_B$ and uses it to update its local density. The IC of the fused density is then given by
\begin{equation}
    I_{\text{fused}} = \text{LCM}(I_A, I_B) = I_A \times I_B / \text{GCD}(I_A, I_B), \label{eq_I_fused}
\end{equation}
where the LCM operation captures all distinct prime factors across both nodes. The shared information between nodes $A$ and $B$ can be determined as
$\text{GCD}(I_A, I_B)$. By factoring out this shared information, the fused density effectively integrates only the new information from node $B$ that is not present in node $A$. The formulation \eqref{eq_I_fused} aligns with the fusion principle in \eqref{eq_ig_fusion}, ensuring that the fused density maintains a complete representation of its contributing data while removing redundancy. Similar to \eqref{eq_multi_fusion}, when fusing information from multiple sources, the fused IC can be computed as
\begin{equation} \textstyle 
    \mathbb{I}_{\text{fused}}(I_1,\dots,I_n) = \prod_{i=1}^n \mathbb{I}_i^{(-1)^{i-1}}, \label{eq_multi_IC_fusion}
\end{equation}
where $\mathbb{I}_i$ represents the combination of $i$ ICs such that
\begin{subequations}
\begin{align}
    \mathbb{I}_1 &= \textstyle \prod_{j=1}^n I_j,  \\
    \mathbb{I}_2 &= \textstyle \prod_{1 \leq j_1 < j_2 \leq n} \text{GCD}(I_{j_1}, I_{j_2}), \cdots  \\
    \mathbb{I}_n &= \textstyle \text{GCD}(I_1, I_2, \dots, I_n). 
\end{align}
\end{subequations}
We note that the alternating multiplication and division of ICs in \eqref{eq_multi_IC_fusion} effectively identifies and removes shared prime factors. As compared to IG-based approaches, PRIMEX significantly reduces communication overhead while still maintaining optimal fusion without knowledge of the IG structure. Examples on PRIMEX effectiveness can be found in \cite[Sec. III]{chang2025primex}.

\begin{remark}
    It is not necessary to explicitly assign prime numbers and compute the GCDs and LCMs. Instead, we can represent each IC using a binary prime exponent vector (PEV) $\phi$, where each entry is either 0 or 1, indicating the absence or presence of a prime factor. With PEVs, the GCD and LCM operations can be performed using element-wise minimum and maximum operations, respectively, significantly reducing complexity and transmission overhead \cite{chang2025primex}.
\end{remark}

\section{Problem Formulation}

We consider distributed single target tracking over a sensor network consisting of communication nodes and sensor nodes. The set of sensor nodes is $\mathcal{S}$, and the set of communication nodes is $\mathcal{C}$, with $\mathcal{N} = \mathcal{S} \cup \mathcal{C} = \{1,\dots,N\}$. All nodes can process and exchange information, but only the sensor nodes can collect measurements. The connectivity among the nodes can be represented using a directed graph $(\mathcal{N},\mathcal{E})$, where $\mathcal{E}\subseteq \mathcal{N}\times\mathcal{N}$ is the set of directed edges. Node $j$ can transmit data to node $i$ if and only $(j,i)\in\mathcal{E}$. In this case, we call node $j$ an in-neighbor of node $i$ and node $i$ an out-neighbor of node $j$. For node $i$, its set of in-neighbors node is $\mathcal{N}^i = \{j \in \mathcal{N}: (j,i) \in \mathcal{E}\}$, which excludes $i$, i.e., there is no self-loop. 

The target state $x_k \in \mathbb{R}^{n_x}$ at time step $k$ evolves over time according to a Markovian motion model
\begin{equation}
    x_{k+1} = f(x_{k}) + q_k, \label{eq_motion}
\end{equation}
where $q_k$ is a zero-mean process noise. Each node $i\in\mathcal{N}$ has a prior density $p_{1\mid0}^i(x_1)$ at time $0$. At time $k = 1,2,\dots$, each sensor node $i \in \mathcal{S}$ collects a measurement $z_k^i \in \mathbb{R}^{n_z}$ according to the measurement model
\begin{equation}
    z_k^i = h^i(x_k) + r_k^i, \label{eq_measurement}
\end{equation}
where $r_k^i$ is a zero-mean measurement noise. The process noise $\{q_k\}$ and measurement noises $\{r_k^i\}$ are mutually independent. 

If there were no communications, the local posterior density $p^i_{k \mid k}(x_k)$ at each sensor node $i$, $i\in\mathcal{S}$, would be given by the following recursive Bayesian filtering equations:
\begin{align}
    p^i_{k \mid k-1}(x_k) &= \int p(x_k \mid x_{k-1}) p^i_{k-1 \mid k-1}(x_{k-1}) dx_{k-1}, \label{eq_pred}\\
    p^i_{k \mid k}(x_k) &\propto \frac{p(z_k^i \mid x_k) p^i_{k \mid k-1}(x_k)}{\int p(z_k^i \mid x_k) p^i_{k \mid k-1}(x_k) d x_k}, \label{eq_upd}
\end{align}
where $p(x_k \mid x_{k-1})$ is the transition density corresponding to the dynamical model \eqref{eq_motion}, and $p(z_k^i \mid x_k)$ is the likelihood function corresponding to the measurement model \eqref{eq_measurement}. The communication nodes only perform the prediction step \eqref{eq_pred}, i.e., $p^i_{k \mid k}(x_k) = p^i_{k \mid k-1}(x_k)~\forall i \in \mathcal{C}$. At the fusion stage, each node $i \in \mathcal{N}$ can leverage information received from its in-neighbors $\mathcal{N}^i$ to improve its local estimate. 

The objective is to design efficient distributed target tracking algorithms that enable each node to recursively compute its local posterior density $p^i_{k\mid k}(x_k)$, by fusing its local information with information received from its in-neighbors, reaching consistent estimates, while minimizing communication overhead.

\section{Proposed Distributed Tracking Algorithms}

In this section, we present two efficient distributed tracking algorithms leveraging PRIMEX: a consensus-based PRIMEX method and a gossip-based PRIMEX method. Both methods incorporate an event-triggered communication mechanism to further enhance communication efficiency.

\subsection{Memoryless Fusion with PRIMEX}


In PRIMEX, each unique prior and sensor measurement at each time step is assigned a distinct prime number. For the allocation of prime numbers without duplication, we refer to the method described in \cite[Sec. II-C]{chang2025primex}. To efficiently assign prime numbers to ICs, we represent ICs using binary PEVs, as described in Section II-B. In PRIMEX, where nodes have local memory, each node stores all its past fusion results (including local densities and their associated ICs) to accurately identify the shared information during fusion. However, in distributed tracking scenarios, the stored densities at previous time steps need to be propagated forward in time for fusion, increasing the computational burden. Therefore, we adopt a memoryless fusion strategy, where each node only maintains its latest local density and IC.

Specifically, each node $i\in\mathcal{N}$ maintains a local posterior density $p^i_{k \mid k}(x_k)$ along with its associated IC $\phi^i_{k \mid k}$ at time step $k$. The ICs remain unchanged during the prediction step \eqref{eq_pred}, i.e., $\phi^i_{k \mid k-1} = \phi^i_{k-1 \mid k-1}$. During the update step \eqref{eq_upd}, if node $i \in \mathcal{S}$ collects a new measurement, it updates its IC by setting the entry corresponding to the assigned prime number to 1.

In the fusion step, when node $i$, with local density $p^i_{k \mid k}(x_k)$ and IC $\phi^i_{k \mid k}$, receives information from an in-neighbor node $j \in \mathcal{N}^i$ with local density $p^j_{k \mid k}(x_k)$ and IC $\phi^j_{k \mid k}$, it fuses the received density $p^j_{k \mid k}(x_k)$ with its local density $p^i_{k \mid k}(x_k)$, leveraging shared information retrieved from the ICs. If the two nodes have the same IC, i.e., $\phi^i_{k \mid k} = \phi^j_{k \mid k}$, then no fusion needs to be performed, and the local density $p^i_{k \mid k}(x_k)$ at node $i$ remains unchanged. When $\phi^i_{k \mid k} \neq \phi^j_{k \mid k}$, the shared information can be identified by computing the element-wise minimum $\min(\phi^i_{k \mid k}, \phi^j_{k \mid k})$ of their PEVs. After incorporating the information of node $j$, node $i$ updates its IC to $\max(\phi^i_{k \mid k}, \phi^j_{k \mid k})$.

Optimal fusion of densities requires the removal of shared information during fusion, as shown in~\eqref{eq_ig_fusion} and~\eqref{eq_multi_fusion}. However, with memoryless fusion, each node only keeps its latest density and IC, making it impossible to reconstruct the density that corresponds exactly to the shared information. One effective solution is to approximate this density as a weighted geometric mean of the two local densities (which is also their weighted Kullback-Leibler average \cite{battistelli2014kullback}). 
Specifically, we let the density that corresponds to the shared information between node $i$ and node $j$ be 
\begin{equation}
    p_{\text{shared}}(x_k)
    = \frac{p^i_{k \mid k}(x_k)^{w_i} p^j_{k \mid k}(x_k)^{w_j}}{\int p^i_{k \mid k}(x_k)^{w_i} p^j_{k \mid k}(x_k)^{w_j} dx_k}, 
    \label{eq_shared_approx}
\end{equation}
with $w_i + w_j = 1$, where the weights $w_i$ and $w_j$ quantify how much of the shared information is attributed to the two local densities. Intuitively, \eqref{eq_shared_approx} takes a convex combination of the log-densities, so it emphasizes regions where both $p^i_{k|k}(\cdot)$ and $p^j_{k|k}(\cdot)$ assign high probability. This makes it a principled proxy for ``shared information''. The fused density at node $i$ then follows the optimal fusion rule,
\begin{equation}
    p_{\text{fused}}(x_k) \propto
    \frac{p^i_{k \mid k}(x_k)\, p^j_{k \mid k}(x_k)}{
    p_{\text{shared}}(x_k)} \propto p^i_{k \mid k}(x_k)^{1-w_i} p^j_{k \mid k}(x_k)^{1-w_j},
    \label{eq_fused_density}
\end{equation}
which has the same functional form as \eqref{eq_shared_approx} and coincides with the generalized covariance intersection (GCI) fusion rule \cite{julier2008fusion}.


The remaining question is how to determine the weights $w_i$ and $w_j$ in a way that reflects the actual overlap of information contributions. This can be achieved by examining the ICs $\phi^i_{k \mid k}$ and $\phi^j_{k \mid k}$, which encode the information pedigree of the local densities. Concretely, the linear combination of their ICs, i.e., $w_i \phi^i_{k \mid k} + w_j \phi^j_{k \mid k}$, should match the IC $\min(\phi^i_{k \mid k}, \phi^j_{k \mid k})$ encoding the shared information as closely as possible. Therefore, we choose the weights by solving the least-squares problem
\begin{equation}
    \underset{w_i,w_j}{\arg\min}
    J(w_i, w_j) 
    = \| w_i\phi^i_{k \mid k} + w_j\phi^j_{k \mid k} 
          - \min(\phi^i_{k \mid k}, \phi^j_{k \mid k}) 
      \|_2^2,
    \label{eq_ls_problem}
\end{equation}
subject to $w_i + w_j = 1$. 

To solve \eqref{eq_ls_problem}, we first stack the weights into $d = [w_i\; w_j]^T$ and define $A = [\phi^i_{k \mid k}\, \phi^j_{k \mid k}]$ and $b = \min(\phi^i_{k \mid k}, \phi^j_{k \mid k})$ such that $J(w_i,w_j) = \|A d - b\|_2^2$. The equality constraint $w_i + w_j = 1$ can be written as $c^T d = 1$ with $c = [1\;1]^T$. The corresponding Lagrangian is
\begin{equation}
    \mathcal{L}(d,\lambda) = \|A d - b\|_2^2 + \lambda (c^T d - 1),
\end{equation}
and the Karush-Kuhn-Tucker (KKT) conditions are
\begin{equation}
        2A^T (A d - b) + \lambda c = 0, \quad
        c^T d - 1 = 0.
\end{equation}
These can be written as the linear system
\begin{equation}
    \begin{bmatrix}
        2A^T A & c \\
        c^T     & 0
    \end{bmatrix}
    \begin{bmatrix}
        d \\ \lambda
    \end{bmatrix}
    =
    \begin{bmatrix}
        2A^T b \\ 1
    \end{bmatrix}.
    \label{eq_kkt_system}
\end{equation}
The weights $(w_i,w_j)$ that define $p_{\text{shared}}(\cdot)$ in \eqref{eq_shared_approx} and thus the fused density $p_{\text{fused}}(\cdot)$ in \eqref{eq_fused_density} can be obtained by solving \eqref{eq_kkt_system}. The least-squares solution of \eqref{eq_kkt_system} minimizes the discrepancy between IC-encoded information, yielding an approximation of the shared information term.

    
    A further practical issue is that the PEV dimension grows with time as more measurements have been received. Specifically, for $|\mathcal{S}|$ sensor nodes at time step $k$, the PEV dimension can be as large as $|\mathcal{S}| \times (k+1)$, where the additional factor of $+1$ accounts for the initial prior at time step $0$. This growth can lead to increased memory and communication overhead when transmitting ICs. To address this, we can restrict the PEVs to a sliding time window \cite{chang2025primex}, so that only information pedigrees from the most recent time steps are tracked. For instance, if the sliding window length is set to $K$, then the memory and communication overhead associated with transmitting ICs is limited to $|\mathcal{S}| \times K$ bits. This mechanism keeps the PEVs at a manageable size and simplifies the least-squares problem used to determine the fusion weights.

It is instructive to highlight the close connection between the proposed PRIMEX-based fusion rule and GCI. In GCI, one forms a weighted geometric mean of the local densities and selects the weights, e.g., by minimizing a scalar function of the fused covariance for Gaussian distributions \cite{julier2008fusion}, without explicitly modeling which parts of the information are actually shared. In our PRIMEX-based approach, the fused density has the same functional form as in GCI, but the weights are no longer free tuning parameters. Instead, they are derived from the overlap of the ICs that encode the information pedigree, via a least-squares fit of the shared information. In this sense, the proposed method can be interpreted as a PRIMEX-guided GCI rule, where the ICs provide structural knowledge about the shared information and the least-squares step translates this structure into the fusion weights.

\subsection{Distributed Tracking Based on Consensus}

We now extend the PRIMEX-based memoryless fusion rule to distributed tracking over a sensor network, represented using graph $(\mathcal{N},\mathcal{E})$. A natural approach is to adopt a consensus-style protocol with a fixed number of communication rounds at each time step. In each round, each node $i \in \mathcal{N}$ updates its local density $p^i_{k \mid k}(x_k)$ and IC $\phi_{k\mid k}^i$ by fusing information sent from their in-neighbors $\mathcal{N}^i$, applying the PRIMEX-based memoryless fusion rule pairwise along the edges.

Specifically, let $p^{i,(l)}_{k\mid k}(x_k)$ and $\phi^{i,(l)}_{k\mid k}$ denote the density and IC at node $i \in \mathcal{N}$ at time step $k$ and consensus round $l$. In a given round, each node $i$ collects the densities and ICs from its in-neighbors $j \in \mathcal{N}(i)$ and fuses them sequentially using the pairwise PRIMEX-based rule described in Section IV-A. For each neighbor $j$, node $i$ interprets $p^{i,(l)}_{k\mid k}(x_k)$ and $p^{j,(l)}_{k\mid k}(x_k)$ as two local densities to be fused, constructs the approximate fused density according to \eqref{eq_fused_density} and \eqref{eq_ls_problem}, and updates its own density to the fused density. The associated ICs are updated by taking the element-wise maximum of the local and the received ICs. After looping over all its in-neighbors, node $i$ obtains an updated density $p^{i,(l+1)}_{k\mid k}(x_k)$ and IC $\phi^{i,(l+1)}_{k\mid k}$. After completing $L$ rounds of consensus fusion, the final density and IC at node $i$ are set as $p^i_{k\mid k}(x_k) = p^{i,(L)}_{k\mid k}(x_k)$ and $\phi^i_{k\mid k} = \phi^{i,(L)}_{k\mid k}$.

Compared with traditional consensus-based distributed fusion algorithms \cite{battistelli2014kullback}, which typically applies GCI with fixed or numerically tuned weights, the PRIMEX-based consensus scheme explicitly exploits the ICs to approximate and remove shared information at each fusion step. This pedigree-aware treatment improves the quality of the consensus updates and, importantly, reveals how much new information each neighbor can provide, which in turn suggests distributed fusion rules that are more accurate and computationally efficient.

\subsection{Distributed Tracking Based on Greedy Gossip}
The IC structure provided by PRIMEX can also be used to move beyond full consensus and design more computationally efficient protocols. An alternative to consensus is gossip-based fusion, where in each communication round, each node only fuses with the information received from a single in-neighbor instead of all its in-neighbors. 

In randomized gossip \cite{ma2016gossip}, the neighbor is chosen uniformly at random, which can substantially reduce the computational load but may result in slow convergence. Greedy gossip \cite{wan2018distributed,shin2020sample} addresses this issue by letting each node inspect all its in-neighbors and select the one that is most different in terms of some discrepancy measure, thereby accelerating convergence at the price of increased computation.

In PRIMEX-based fusion framework, greedy gossip can be implemented in a way that directly exploits the information pedigree. Specifically, at communication round $l$ and time step $k$, node $i \in \mathcal{N}$ holds a local density $p^{i,(l)}_{k\mid k}(x_k)$ and an IC $\phi_{k\mid k}^{i,(l)}$. For each in-neighbor $j \in \mathcal{N}^i$, node $i$ evaluates how much new information node $j$ can provide by computing the IC increment
\begin{equation}
    \Delta\phi^{j\rightarrow i,(l)}_{k\mid k}
= (\phi^{j,(l)}_{k\mid k} - \phi^{i,(l)}_{k\mid k})_+ ,
\end{equation}
where the positive part is taken element-wise. The scalar score 
\begin{equation}
    s_{ij}^{(l)} = \|\Delta\phi^{j\rightarrow i,(l)}_{k\mid k}\|_1,
\end{equation}
then qualities the amount of fresh information that node $j$ can offer to node $i$, in terms of the number of bits in the PEV representation of IC present at node $j$ but not yet at node $i$. In a greedy gossip step, node $i$ selects the in-neighbor 
\begin{equation} \textstyle 
    j^\star = \arg\max_{j \in \mathcal{N}(i)} s^{(l)}_{ij},
\end{equation}
and performs a pairwise PRIMEX-based memoryless fusion with node $j^\star$ according to \eqref{eq_fused_density} and \eqref{eq_ls_problem}. The IC $\phi^{i,(l+1)}_{k \mid k}$ at node $i$ is set to $\max(\phi^{i,(l)}_{k \mid k}, \phi^{j,(l)}_{k \mid k})$, ensuring that the pedigree continues to track which information has been disseminated. After completing $L$ rounds of greedy gossip-based fusion, the final density and IC at node $i$ are set as $p^i_{k\mid k}(x_k) = p^{i,(L)}_{k\mid k}(x_k)$ and $\phi^i_{k\mid k} = \phi^{i,(L)}_{k\mid k}$.

This design has two attractive features. First, the informativeness of an in-neighbor is evaluated purely in the IC domain, using cheap element-wise operations on binary vectors, so the neighbor selection step is lightweight. Second, the selection criterion is tightly coupled to the actual information flow. By maximizing the new information acquired at each interaction, the greedy gossip process systematically prioritizes edges that contribute fresh measurements, leading to faster information spread over the sensor network while maintaining the computation savings of gossip relative to full consensus. 

In connected networks, greedy gossip based PRIMEX empirically converges to the same information pedigree as consensus based PRIMEX, though potentially at different mixing rates. This has been verified in our simulations in Section V.

\subsection{Event-Triggered PRIMEX Fusion}

Both the consensus-based and gossip-based fusion schemes described above assume that nodes exchange information in every communication round. While this simplifies the protocol, it can lead to unnecessary transmissions when a node's local information has not sufficiently changed. Hence, communication efficiency can be further improved by making data transmission event-triggered rather than ``always-on''. Specifically, a node transmits information to its out-neighbors only when its local information has changed sufficiently relative to its last transmission \cite{dong2022event}. 

In PRIMEX-based fusion, this decision can be made directly by exploiting structural changes in the ICs, instead of monitoring numerical changes in the local densities. In the first communication round $l=1$, each node $i\in\mathcal{N}$ maintains a reference IC $\bar\phi^i_{k\mid k}$ corresponding to its last transmission before fusion. At the beginning of subsequent rounds $l \geq 2$, node $i$ checks whether any new information has been accumulated since its last transmission by comparing $\phi^{i,(l-1)}_{k\mid k}$ with $\bar\phi^i_{k\mid k}$. If at least one element has changed from 0 to 1, then node $i$ has acquired new information that were not previously communicated and should transmit its current density and IC to its out-neighbors. Otherwise, node $i$ skips transmission in round $l$, but can still receive and fuse information from neighbors that do transmit. After a transmission, the reference is updated as $\bar\phi^i_{k\mid k} \leftarrow \phi^{i,(l-1)}_{k\mid k}$. In this way, PRIMEX-based fusion naturally supports a simple event-triggered communication policy driven by structural changes in the information pedigree.

An important advantage of this IC-based trigger is that, for consensus-style PRIMEX, it does not change the information that ultimately propagates through the network. The pairwise PRIMEX fusion rule only updates a node when the two ICs differ. The trigger follows the same logic at the communication level, since a node transmits only when it has ever gained new information, in terms of IC, relative to its last transmission. In a connected network, every new information is still transmitted through the network until all the nodes have incorporated it, regardless of whether redundant transmissions are suppressed. Once all nodes share the same IC, further transmissions would not change any estimates. The trigger therefore preserves the same final fused information under connected networks and sufficient communication rounds.

In much of the event-triggered distributed fusion literature, each node needs to store a reference density corresponding to its last transmission to evaluate whether to transmit again. The decision is made by comparing the current local density with the reference density, and the triggering condition is typically based on a distance measure, such as the Mahalanobis distance \cite{ge2019distributed} or the Kullback-Leibler divergence \cite{dong2022event}. In addition to the storage overhead of maintaining the reference density, evaluating these distance measures can be computationally intensive and numerically sensitive, especially in high-dimensional state spaces. As a comparison, 
the PRIMEX-based trigger operates purely in the IC domain, where the discrepancy is measured arithmetically with binary PEVs. Therefore, the triggering rule is both lightweight and robust, making event-triggered communication particularly well-suited in PRIMEX-based fusion.

\section{Simulations and Results}

We evaluate the performance of the two proposed distributed tracking algorithms and compare them with centralized fusion and distributed tracking based on covariance intersection in a simulation study.

\begin{figure}[!t]
    \centering
    \includegraphics[width=0.95\columnwidth]{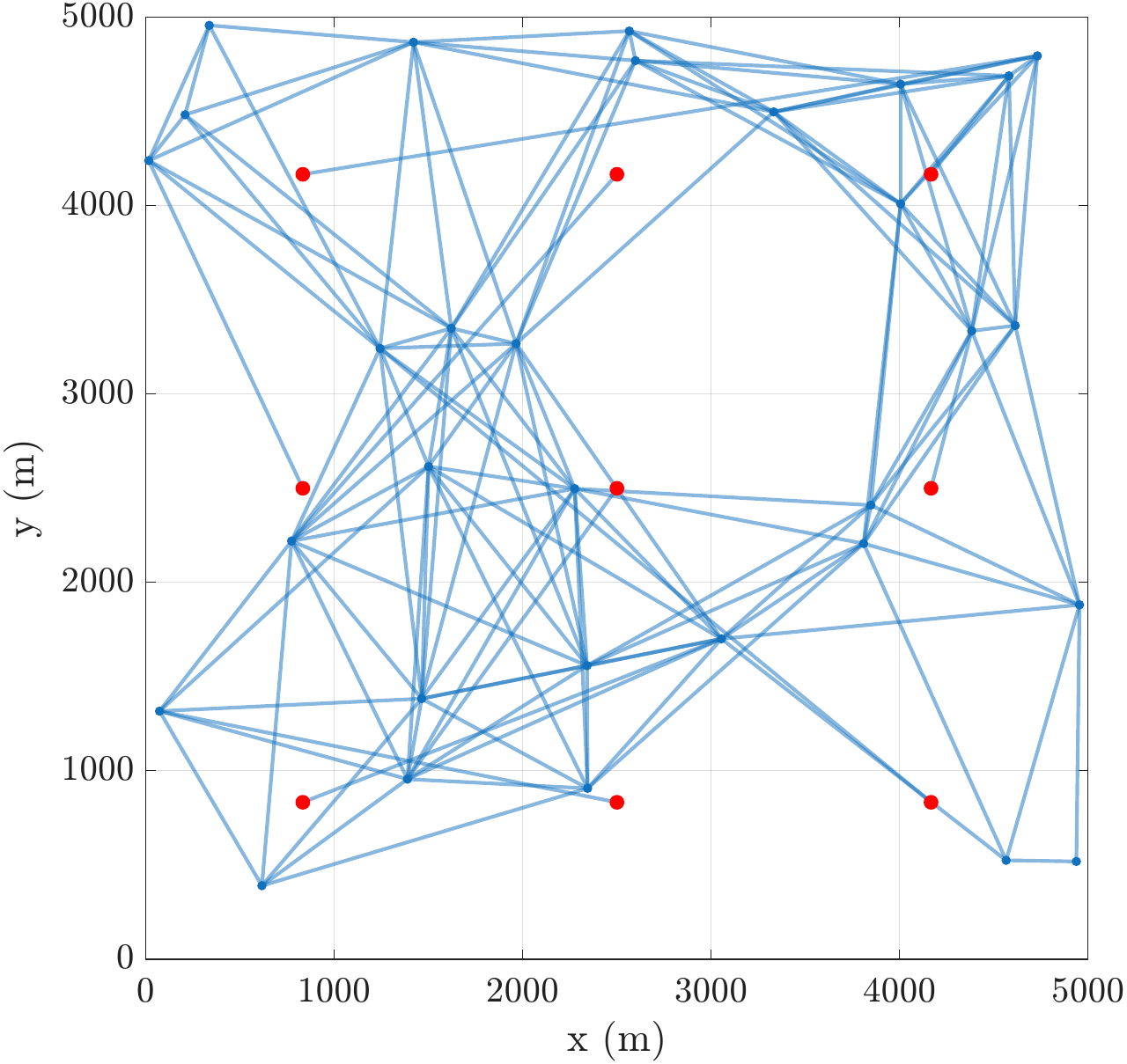}%
    \caption{Sensor network topology. Red dots denote sensor nodes and blue dots denote communication nodes. Edges indicate communication links, which are assumed to be bidirectional. This illustration emphasizes connectivity only; the connectivity between nodes is independent of their physical distances.}%
    \label{fig_sensor_network}%
\end{figure}

We consider a 2D scenario with $K= 50$ time steps in total, where the target moves according to a nearly constant velocity model with state vector $x_k = [p^x_{k}\,v^{x}_k\,p^{y}_k\,v^y_{k}]^T$, consisting of 2D position $(p^x_{k},p^y_{k})$ and 2D velocities $(v^x_{k},v^y_{k})$. The motion model \eqref{eq_motion} is $f(x_k) = F x_k + q_k$, with transition matrix
\begin{equation}
    F = \begin{bmatrix}
        1 & T \\
        0 & 1 \\
    \end{bmatrix} \otimes I_2,
\end{equation}
where $T = 1\,\textrm{s}$ is the sampling time, $\otimes$ denotes the Kronecker product, and $I_2$ is the $2 \times 2$ identity matrix. The process noise $q_k$ is zero-mean Gaussian with covariance
\begin{equation}
    Q = 25 \begin{bmatrix}
        T^3/3 & T^2/2 \\
        T^2/2 & T \\
    \end{bmatrix} \otimes I_2.
\end{equation}
The sensor network consists of $N=40$ nodes, including $9$ sensor nodes and $31$ communication nodes, as shown in Fig.~\ref{fig_sensor_network}. Each node on average has 6.85 neighbors. Each sensor node uses the same measurement model \eqref{eq_measurement} with $z_k = H x_k + r_k$, where $H$ is the observation matrix and the sensor noise $r_k$ is zero-mean Gaussian with covariance $R = 100I_2$. The target has a global prior density $p_{1\mid 0}(x_k)$ at time $0$ shared by all the nodes, which is Gaussian with mean $[0\,100\,0\,100]^T$ and covariance $25I_4$. The target trajectory is generated by sampling from the motion model, starting from the prior mean.

We compare the following six different algorithms:
\begin{itemize}
    \item Centralized fusion (CF): All sensor nodes transmit their measurements to a fusion center, which performs centralized estimation using an information filter. As an optimal fusion approach, CF serves as a performance benchmark for distributed algorithms.
    \item Covariance intersection (CI): Each node performs fusion using CI within a consensus-based protocol. We consider two variants. For the first variant, in each communication round, every node fuses its local density with the densities received from all its in-neighbors using CI with uniform weights. We refer to this method to as CI with uniform weights (CI-UW). For the second variant, in each communication round, each node sequentially fuses its local density with the densities received from its in-neighbors using CI, where the fusion weight in each step is chosen by minimizing the trace of the resulting fused covariance matrix \cite{deng2012sequential}. We refer to this method as CI with optimized weights (CI-OW).
    \item The consensus-based PRIMEX with event-triggered communication is referred to as {PRIMEX-C-ET}. Note that its always-on counterpart (without event-triggered communication) achieves the same estimation performance. Hence, it is sufficient to only report the results of PRIMEX-C-ET.
    \item The greedy gossip-based PRIMEX with event-triggered communication is referred to as {PRIMEX-G-ET}. It shows similar estimation performance as its always-on counterpart, so we only report the results of PRIMEX-G-ET.
\end{itemize}


Estimation performance is quantified using the Root Mean Square Error (RMSE) of the 2D position estimates, averaged over all nodes. Let $x_k$ denote the true state at time step $k$, and $\hat{x}^i_k$ the local estimate at node $i \in \mathcal{N}$. The network-averaged position RMSE at time $k$ is
\begin{equation} \textstyle 
\mathrm{RMSE}_{k}
=
\sqrt{
  \frac{1}{|\mathcal{N}|}
  \sum_{i \in \mathcal{N}}
  \|
    H x^i_k - H x_k
  \|^2
},
\end{equation}
and the reported RMSE is obtained by averaging $\mathrm{RMSE}_{k}$ over all time steps. Computational efficiency is assessed using the average runtime per time step. The communication efficiency of PRIMEX-C-ET and PRIMEX-G-ET is evaluated via transmission rate, defined as the average number of transmissions per node, per communication round, and per time step. All the results are obtained by averaging over 100 Monte Carlo runs.


\begin{figure}[!t]
    \centering
    \includegraphics[width=\columnwidth]{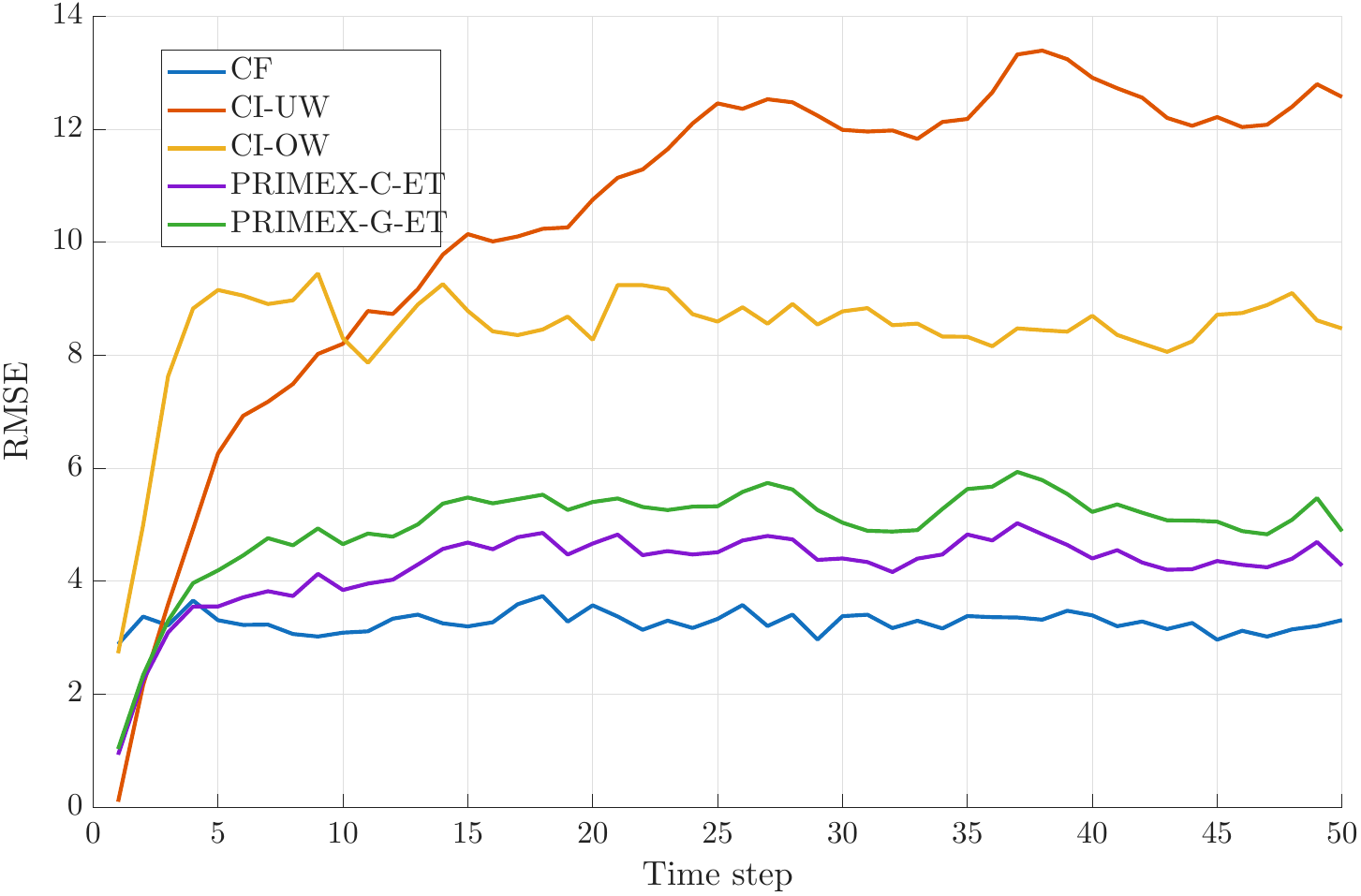}%
    \caption{RMSE over time for different algorithms. For distributed algorithms, the number of communication rounds is set to 7.}%
    \label{fig_rmse_over_time}%
\end{figure}

\begin{figure}[!t]
    \centering
    \includegraphics[width=0.9\columnwidth]{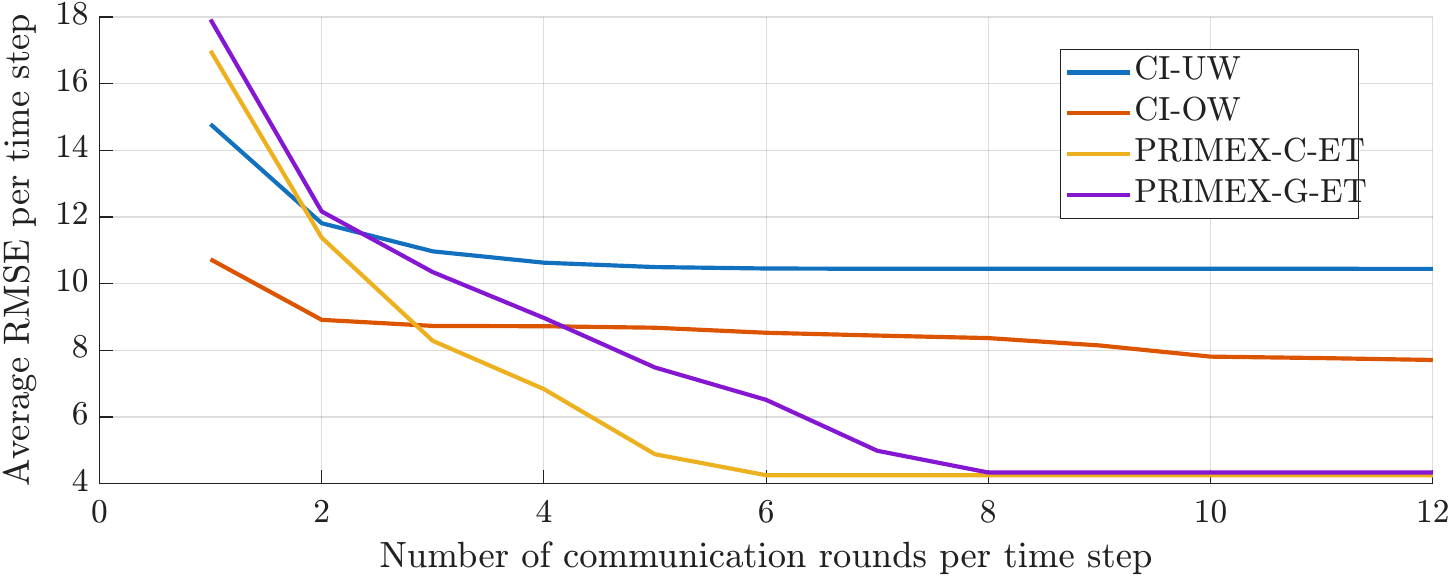}%
    \caption{Average RMSE per time step versus the number of communication rounds for different distributed tracking algorithms.}%
    \label{fig_rmse_vs_round}%
\end{figure}

\begin{figure}[!t]
    \centering
    \includegraphics[width=\columnwidth]{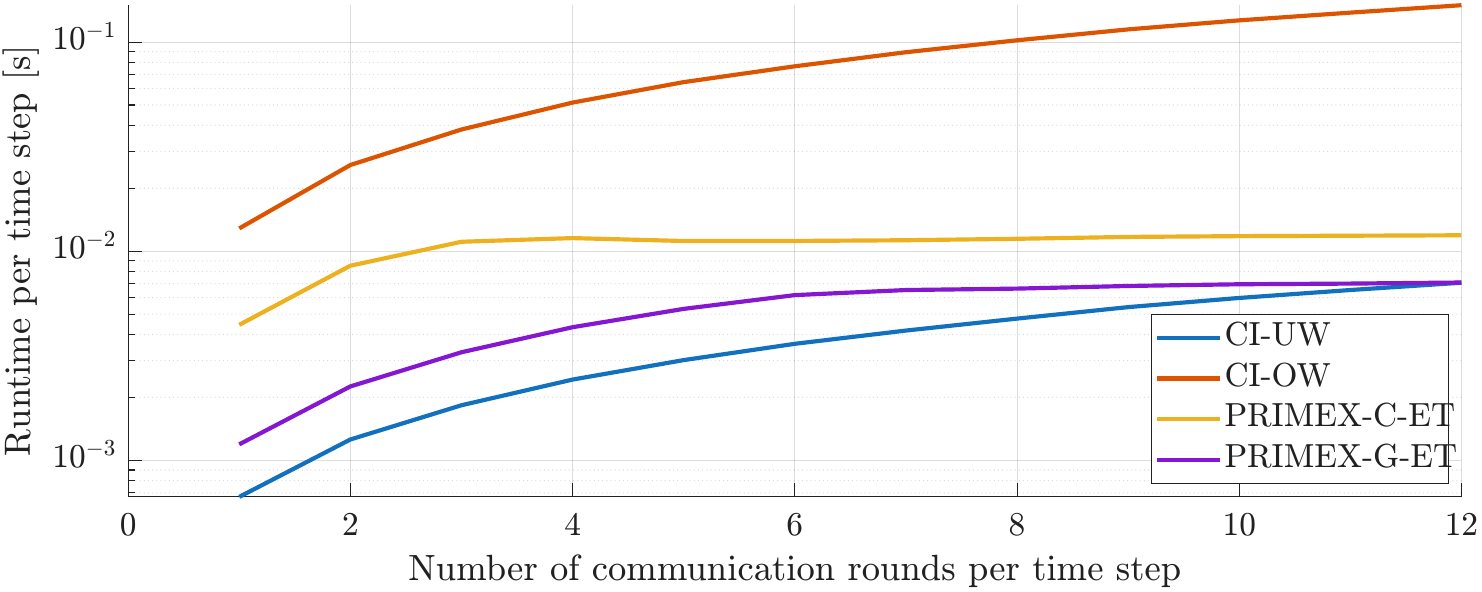}%
    \caption{Average runtime per time step versus the number of communication rounds for different distributed tracking algorithms. All algorithms are implemented in MATLAB R2025a on a MacBook Air with an M4 chip.}%
    \label{fig_runtime_vs_round}%
\end{figure}

\begin{figure}[!t]
    \centering
    \includegraphics[width=\columnwidth]{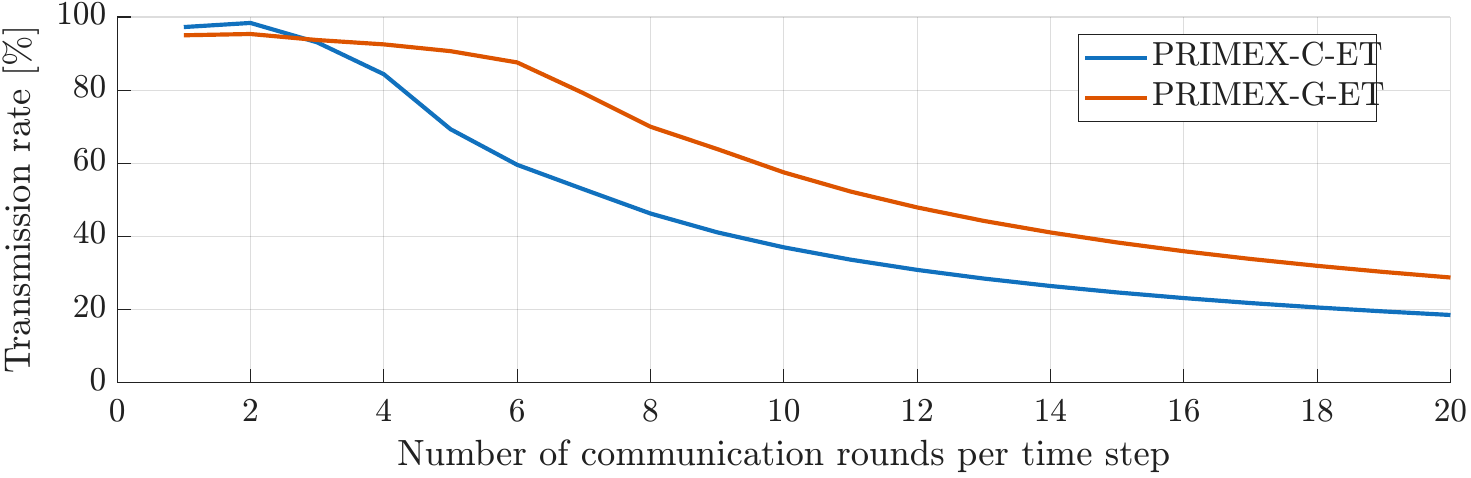}%
    \caption{Transmission rate (in percentage) versus the number of communication rounds for PRIMEX-C-ET and PRIMEX-G-ET.}%
    \label{fig_transmission_rate}%
\end{figure}

Fig. \ref{fig_rmse_over_time} presents the network-averaged RMSE over time for the different algorithms, with the number of communication rounds for the distributed algorithms fixed to 7. As expected, CF provides the ideal benchmark with the lowest estimation error. Both PRIMEX-based methods achieve performance that stays close to CF, while outperforming both CI-UW and CI-OW by a large margin. This confirms that, despite operating in a fully distributed manner, the PRIMEX-based algorithms can  approach centralized performance by exploiting the information pedigree encoded in the ICs.

Fig. \ref{fig_rmse_vs_round} reports the average RMSE per time step as a function of the number of communication rounds for the distributed methods. As the number of communication rounds increases, all methods have improved estimation accuracy, reflecting the benefit of additional information exchange. For a very small number of rounds, the PRIMEX fusion rule operates with only partially mixed PEVs, so the least-squares approximation of the shared information can be relatively crude; in this regime, the conservative CI fusion rule can yield slightly lower RMSE. As the number of rounds increases, however, the information pedigrees across nodes become more aligned and the least-squares approximation of the shared part improves, allowing PRIMEX to control double counting more accurately than CI and thereby achieve better estimation performance. In addition, once the network is sufficiently mixed (with a moderate or large number of communication rounds), PRIMEX-C-ET and PRIMEX-G-ET exhibit almost identical performance.

Fig. \ref{fig_runtime_vs_round} illustrates the computational cost, measured as average runtime (in seconds) per time step, versus the number of communication rounds on a logarithmic scale. As expected, the runtime of all distributed algorithms increases with the number of rounds, since more rounds imply more local fusions per time step. CI-UW has the lowest computational cost per round due to its simple uniform weight update, whereas CI-OW is more expensive because it solves an optimization problem at each pairwise fusion. The PRIMEX-based methods fall between these two: they incur additional cost from solving the least-squares problem in the IC domain. It is also noticeable that, when the network has become well mixed, the runtime of PRIMEX-C-ET and PRIMEX-G-ET grows only marginally with additional rounds. In this regime, all nodes have nearly identical ICs, so they rarely transmit or perform fusions.

Finally, Fig. \ref{fig_transmission_rate} reports the transmission rate for PRIMEX-C-ET and PRIMEX-G-ET. In general, for both methods, the transmission rate decreases as the number of communication rounds increases, since the network becomes better mixed and the event-trigger mechanism suppresses redundant transmissions. PRIMEX-C-ET typically operates at a lower transmission rate than PRIMEX-G-ET, reflecting the slower mixing under the greedy gossip based communication protocol, where each node can incorporate information from at most one neighbor per round. However, as seen in Fig. \ref{fig_runtime_vs_round}, this also leads to lower runtime for PRIMEX-G-ET compared with PRIMEX-C-ET, illustrating a trade-off between communication efficiency and computational burden for the two fusion methods.

\section{Conclusions and Future Work}

In this paper, we have proposed PRIMEX-based distributed tracking algorithms. Using the PEVs to encode the information pedigree, we have presented a memoryless PRIMEX fusion strategy that approximates the shared information via a least-squares fit in the IC domain. We have also embedded this rule into two distributed architectures, a consensus-based PRIMEX and a greedy gossip-based PRIMEX, and then combined them with event-triggered communication. Simulation results show that the proposed PRIMEX-based methods outperform CI by a significant margin and substantially reduce communication load and computational burden.

A few directions remain for future work. The first concerns richer event-triggered communication strategies. In this paper, a node transmits whenever its PEV gains any new positive entries compared with its last transmission. This can be refined by requiring that the number of newly activated PEV entries exceeds a threshold before transmission, thereby ignoring very small structural updates. The PEV-based triggers can also be combined with numerical criteria, such as Kullback-Leibler divergence, to react only when both the information pedigree and the local density (or one of them) change sufficiently.

The second direction is extension to heterogeneous sensor networks with strongly different measurement noise covariances. Our current least-squares formulation in the IC domain treats all PEV entries symmetrically. When noise levels vary widely, this may underexploit the reliability differences across sensors. A possible remedy is a weighted least-squares formulation, where each PEV entry associated with a measurement is scaled by a reliability weight, e.g., a function of the inverse trace of its measurement noise covariance \cite{zheng2024distributed}. This would allow information from more accurate sensors to have greater influence when approximating shared information.

\bibliographystyle{IEEEtran}
\bibliography{mybibli.bib}

\end{document}